\documentclass[lettersize,journal]{IEEEtran}
\usepackage{amsmath,amsfonts}
\usepackage{algorithmic}
\usepackage{algorithm}
\usepackage{array}
\usepackage{textcomp}
\usepackage{stfloats}
\usepackage{url}
\usepackage{verbatim}
\usepackage{graphicx}
\usepackage{subcaption}
\usepackage{caption}
\usepackage{cite}
\usepackage{bm}
\usepackage{amsthm,amsmath,amssymb}
\usepackage{mathrsfs}
\usepackage{soul,color,xcolor}
\usepackage{graphicx}
\usepackage{epstopdf}
\def\BibTeX{{\rm B\kern-.05em{\sc i\kern-.025em b}\kern-.08em
    T\kern-.1667em\lower.7ex\hbox{E}\kern-.125emX}}
\usepackage{balance}
\begin{document}
\title{Pinching-Antenna Assisted Simultaneous Wireless Information and Power Transfer}
\author{Yixuan Li, Ji Wang, \IEEEmembership{Senior Member,~IEEE}, Yuanwei Liu, \IEEEmembership{Fellow,~IEEE}, and Zhiguo Ding, \IEEEmembership{Fellow,~IEEE}
\thanks{\textit{Corresponding author: Ji Wang.}}
\thanks{Yixuan Li and Ji Wang are with the Department of Electronics and Information Engineering, Central China Normal University, Wuhan 430079, China (e-mail: yixuanli@mails.ccnu.edu.cn; jiwang@ccnu.edu.cn ).

Yuanwei Liu is with the Department of Electrical and Electronic Engineering, The University of Hong Kong, Hong Kong, China (e-mail: yuanwei@hku.hk).

Zhiguo Ding is with the Department of Computer and Information Engineering, Khalifa University, Abu Dhabi, UAE, and is with the Department of Electrical and Electronic Engineering, the University of Manchester, Manchester, UK, M1 9BB (e-mail: zhiguo.ding@ieee.org).
}}

\maketitle

\begin{abstract}
This letter introduces a novel pinching-antenna-system (PASS) assisted simultaneous wireless information and power transfer (SWIPT), where multiple pinching antennas (PAs) are strategically activiated on a waveguide to facilitate information transmission to multiple information receivers (IRs) and power transfer to multiple energy receivers (ERs) simultaneously. Leveraging the single-waveguide architecture, non-orthogonal multiple access is employed to enable the superposed transmission of information signals, eliminating the need for dedicated energy carriers by concurrently serving both the IRs and the ERs. In this letter, an optimization problem is first formulated with an objective is to maximize the sum power received at the ERs via joint optimizing the IRs power allocation and the PAs positioning. Given the challenging non-convex nature of the formulated optimization problem, it is decoupled into two sub-problems which are solved alternatively. Specifically, the power allocation subproblem is recast a convex optimization problem, and hence can be solved efficiently. Furthermore, we propose a high-precision element-wise algorithm and a low-complexity linearly decreasing weight particle swarm optimization algorithm to solve the position optimization sub-problem. The numerical results demonstrate that PAs assisted SWIPT can achieve a remarkable performance gain compared to conventional system.
\end{abstract}

\begin{IEEEkeywords}
Pinching-antenna systems, pinching beamforming, position optimization, simultaneous wireless information and power transfer.
\end{IEEEkeywords}

\section{Introduction}

With the increasing demand for the energy-constrained Internet of Things (IoT), the simultaneous wireless information and power transfer (SWIPT) technology, by integrating energy and data transmission, has opened up a highly efficient path for its development. In order to maximize the performance of the SWIPT systems within the limited spectrum resources, multi-antenna and flexible-antenna technologies have been developed to reconfigure wireless channels to facilitating signal transmission, such as reconfigurable intelligent surfaces (RISs) \cite{9424177}, movable-antenna systems\cite{10286328}, and fluid-antenna systems \cite{10753482}. However, these technologies are not capable to combat line-of-sight blockage, e.g., movable antennas move the antennas within a range of several wavelengths. Although they can adapt the small-scale fading of the channels, it is difficult for them to mitigate large-scale path-loss of the channels \cite{9424177,10286328}. Moreover, the implementation of these technologies is usually complex due to their intricate hardware designs.

Recently, pinching-antenna systems (PASS) have been developed to provide a new perspective for the development of multi-antenna technologies due to its advantages of the excellent spatial degree of freedom (DoF), low cost, and low manufacturing complexity \cite{10896748} and \cite{10945421} here. Specifically, by deploying pinching antennas (PAs) on a waveguide connected to the base station (BS), the capability of receiving or transmitting signals can be achieved \cite{liu2025pinching,wang2025modeling}. Moreover, by adjusting the position of the PAs, the wireless channels between the PAs and the devices can be sophisticated reconfigured. Different from existing flexible-antenna technologies, a PA can be deployed across the entire waveguide, which means that PASS cannot only adapt small-scale fading of the channels but also mitigate large-scale path-loss. Therefore, PASS is expected to become one of the key technologies for the next-generation wireless communications. Those existing works related to PASS had fully demonstrated the remarkable performance of PASS in wireless communications \cite{10896748,liu2025pinching,wang2025modeling,10945421,10909665,10912473}. In \cite{10896748}, the authors first studied the PA location optimization problem and presented an upper bound on the downlink transmission performance of the PASS with a single user and a single waveguide. The authors of \cite{10909665} investigated the trade-off problem of maximizing the minimum user rate in the downlink transmission of the PASS. The authors of \cite{10912473} investigated a practical implementation of PASS, which can improve the throughput by intelligently activating preinstalled PAs, instead of moving them in real time. The existing works only demonstrated the remarkable ability of the PASS to significantly enhance information transmission. Whether it also has a promoting effect in improving wireless power transfer is worthy of investigation, which motivates this letter.

To the best knowledge of the authors, this is the first work to investigate an SWIPT enabled by PASS, in which multiple PAs are activated on a waveguide to provide information transmission for multiple IRs and power transfer for multiple ERs, respectively. We adopt the non-orthogonal multiple access (NOMA) technique to achieve the superimposed transmission of information signals. To fully explore the potential of the system, we formulate a sum receive power maximization problem through joint optimization of power allocation strategies for the IRs and the ERs, along with positioning optimization for the PAs. For the non-convex original problem with coupled variables, a decomposition strategy is proposed, which decouples the formulated problem into two subproblems, termed power allocation optimization and position optimization, respectively. Specifically, the power allocation subproblem is transformed into a convex optimization form which can be solved via standard convex programming techniques. For position optimization, a high-performance element-wise optimization algorithm and a low-complexity linearly decreasing weight particle swarm optimization (LDW-PSO) algorithm are developed to adapt to various application scenarios. Simulation results validate the superior performance of the proposed system compared to conventional fixed-antenna system.


\section{System Model and Problem Formulation}
\begin{figure}[!t]
\centering
\includegraphics[width=3.0in]{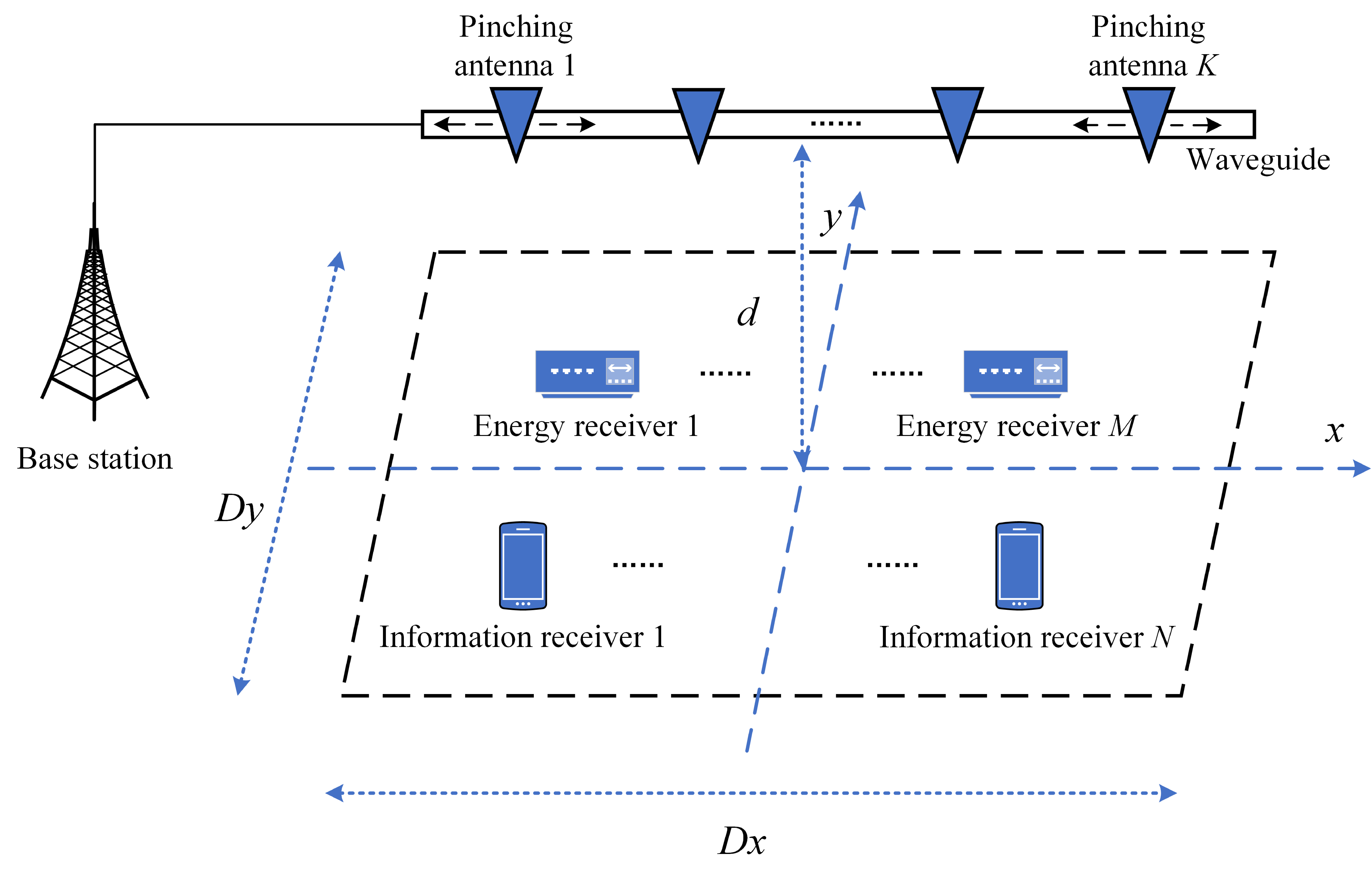}
\caption{A model of SWIPT enabled by PASS.}
\label{system_model}
\end{figure}
As illustrated in Fig. {\ref{system_model}}, we consider an SWIPT system enabled by PASS, which consists of $M$ PAs on one waveguide, $K_E$ single-antenna energy receivers (ERs), and $K_I$ single-antenna information receivers (IRs). For the convenience of mathematical expression, we set ${\cal M} = \left\{ {1, \cdots ,M} \right\}$, ${\cal K_E} = \left\{ {1, \cdots ,K_E} \right\}$, and ${\cal K_I} = \left\{ {1, \cdots ,K_I} \right\}$. In this system, the BS transmits the superimposed information signal along the waveguide, and sends it to the IRs and ERs by utilizing the PAs on the waveguide. Clearly, there is no need to transmit dedicated energy signals over a single waveguide, as information signals can be used to supply energy to the ERs. Each PA can be activated at any position within the range of the entire waveguide. The three-dimensional (3D) coordinate of the $m$-th PA can be represented as ${\bm \psi} _m^{{\rm{PA}}} = \left[ {x_m^{{\rm{PA}}},0,d} \right]$, where $m \in {\cal M}$. The height and the length of the waveguide are denoted by $d$ and $D_w$, respectively. The IRs and the ERs are distributed within a rectangular area with side lengths $D_x$ and $D_y$ respectively. Thus, the 3D coordinates of the $j$-th ER and the $i$-th IR are respectively represented by ${\bm \psi} _j^{{\rm{ER}}} = \left[ {x_j^{{\rm{ER}}},y_j^{{\rm{ER}}},0} \right]$ and ${\bm \psi} _i^{{\rm{IR}}} = \left[ {x_i^{{\rm{IR}}},y_i^{{\rm{IR}}},0} \right]$, where $j \in {\cal K_E}$ and $i \in {\cal K_I}$. The channels of the PASS are based on the near-field model \cite{10896748}, thereby the channels from the $m$-th PA to the $j$-th ER and the $i$-th IR are expressed as
\begin{flalign}
{{h}_{{\rm{E}},j,m}} = {{\frac{{{\eta }{e^{ - j\frac{{2\pi }}{\lambda }\left\| {{\bm \psi} _j^{{\rm{ER}}} - {\bm \psi} _m^{{\rm{PA}}}} \right\|}   }}}{{\left\| {{\bm \psi} _j^{{\rm{ER}}} - {\bm \psi} _m^{{\rm{PA}}}} \right\|}}} \cdot {e^{ - j\frac{{2\pi }}{{{\lambda _g}}}\left\| {{\bm \psi} _0^{\rm PA} - {\bm \psi} _m^{\rm PA}} \right\|}}}, \\
{{h}_{{\rm{I}},i,m}} = {{\frac{{{\eta }{e^{ - j\frac{{2\pi }}{\lambda }\left\| {{\bm \psi} _i^{{\rm{IR}}} - {\bm \psi} _m^{{\rm{PA}}}} \right\|}  }}}{{\left\| {{\bm \psi} _i^{{\rm{IR}}} - {\bm \psi} _m^{{\rm{PA}}}} \right\|}}} \cdot {e^{ - j\frac{{2\pi }}{{{\lambda _g}}}\left\| {{\bm \psi} _0^{\rm PA} - {\bm \psi} _m^{\rm PA}} \right\|}}},
\end{flalign}
respectively, where $\eta  = \frac{c}{{4\pi {f_c}}}$. $\lambda$, $c$, and $f_c$ represent the wavelength, propagation speed, and carrier frequency, respectively. ${\lambda _g} = \frac{\lambda }{{{n_{\rm neff}}}}$ is the guided wavelength in the waveguide. ${\bm \psi}_0^{\rm PA} = \left[ {x_0^{\rm PA},0,d} \right]$ represents the coordinate of the feed point of the waveguide. Since all the PAs are on the same waveguide, the signals need to be superimposed before being passed on the waveguide. {Thus, the transmitted superimposed signal at the BS can be expressed as $x = \sum\nolimits_{i \in {\cal K_I}} {\sqrt {{\overline p}_{i}^{{\rm{IR}}}} c_i^{{\rm{IR}}}}$, where ${\overline p}_{i}^{\rm IR}$ is the signal power for the $i$-th IR, and the power budget is given by $\sum\nolimits_{i \in {\cal K_I}} {{\overline p}_{i}^{\rm IR}}  \le {P_B}$. $P_B$ is maximum transmit power at the BS. $c_i^{\rm IR}$ is the baseband signal sent from the BS to the $i$-th IR, which is modeled as follows: $c_i^{{\rm{IR}}} \sim {\cal CN}\left( {0,1} \right)$.}

For a multi-antenna BS, the flexible and independent transmit power regulation can be achieved by controlling the radio frequency chains and the power amplifiers connected to each antenna. On the contrary, the transmit power on the each PA is related to its coupling length with the waveguide \cite{liu2025pinching}. {In this work, we adopt the equal power model \cite{liu2025pinching} in the PASS. Signal propagation through the waveguide incurs negligible power loss, such that it has a negligible impact on the overall system performance \cite{liu2025pinching}}. In addition, the impact of the additive noise power on the power received at the ERs can also be neglected. Thus, the received power at the $j$-th ER can be expressed as
\begin{flalign}
{E_j} = \sum\nolimits_{i = 1}^{K_I} {{p_{i}^{\rm IR}}{{\left| {\sum\nolimits_{m = 1}^M {h_{{\rm E},j,m}}} \right|}^2}} ,
\end{flalign}
where ${p_{i}^{\rm IR}} = {{\overline p}_{i}^{\rm IR} / M}$ represents the total transmit power at the $m$-th PA for the $i$-th IR. Since all the signals passed on the waveguide are superimposed together, each IR can adopt the successive interference cancellation (SIC) technique to effectively mitigate the impact of the interference. {Without loss of generality, the decoding order of SIC is determined by the sum of the channel gains from all the PAs to each IR, leading to the following constraints
\begin{flalign}
{{{\left| {\sum\nolimits_{m \in {\cal M}} {h_{{\rm I},{1},m}}} \right|}^2}}  \le ... \le {{{\left| {\sum\nolimits_{m \in {\cal M}} {h_{{\rm I},{K_I},m}}} \right|}^2}} , \label{SIC}
\end{flalign}
where ${{{\big| {\sum\nolimits_{m \in {\cal M}} {h_{{\rm I},1,m}}} \big|}^2}}$ denotes the aggregated channel gain from all the PAs to the first decoded IR in the SIC order.}

Furthermore, the received signal-to-interference-plus-noise ratio (SINR) of the $i$-th IR can be expressed as
\begin{flalign}
\!\!\!{{\gamma}_i} = {\frac{{p_{i}^{\rm IR}{{\big| {\sum\nolimits_{m = 1}^M {h_{{\rm I},i,m}}} \big|}^2}}}{{\sum\nolimits_{l = i+1}^{K_I} {p_{l,m}^{\rm IR}{{\big| {\sum\nolimits_{m = 1}^M {h_{{\rm I},i,m}}} \big|}^2}} + { \sigma _i^2}}}},
\end{flalign}
where $\sigma _i^2$ is the noise power received at the $i$-th IR.

We aim to maximize the total received power at the ERs with the constraint of the quality of information transmission. Let ${\bf{x}} = \left\{ {x_1^{\rm PA},...,x_M^{\rm PA}} \right\}$ and ${{\bf{p}}^{\rm IR}} = \left\{ {p_1^{\rm IR},...,p_{K_I}^{\rm IR}} \right\}$. Therefore, the sum receive power maximization problem can be formulated as
\begin{subequations}\label{P1}
\begin{eqnarray}
&\mathop {\max }\limits_{{\bf x},{{\bf p}^{\rm IR}}} &{\sum\nolimits_{j = 1}^{K_E} {{E_j}}} \label{P1-a}\\
&{\rm s.t.}&{{\gamma}_i} \ge {{\gamma}_{\min }},\forall i \in {\cal K_I}, \label{P1-b}\\
&&{E_j} \ge {E_{\min }},\forall j \in {\cal K_E}, \label{P1-c}\\
&&\sum\nolimits_{i \in {\cal K_I}} {{p}_{i}^{\rm IR} M}  \le {P_B},~~~~~ \label{P1-d}\\
&&\left| {x_m^{\rm PA} - x_{m'}^{\rm PA}} \right| \ge \Delta , m' \ne m, \forall m, m' \in {\cal M}, \label{P1-e}\\
&&{x_m^{\rm PA}} \in \left[ 0,D_w \right], \forall m \in {\cal M}, \label{P1-f}\\
&&{{{{\big| {\sum\nolimits_{m \in {\cal M}} {h_{{\rm I},1,m}}} \big|}^2}}  \le ... \le {{{\big| {\sum\nolimits_{m \in {\cal M}} {h_{{\rm I},{K_I},m}}} \big|}^2}}.~~~~~~} \label{P1-g}
\end{eqnarray}
\end{subequations}
where constraints (\ref{P1-b}) and (\ref{P1-c}) respectively indicate that the SINR of each IR and the received power of each ER must be higher than the minimum target values ${\gamma}_{\min}$ and $E_{\min}$. Constraint (\ref{P1-d}) ensures that the upper bound of the transmit power of the BS must not be exceeded. Constraint (\ref{P1-e}) is to avoid antenna coupling, where $\Delta$ is the minimum spacing between any two PAs. Constraint (\ref{P1-f}) indicates that PAs should be deployed within a certain range. {(\ref{P1-g}) is the decoding order constraint for SIC.} It should be noted that problem (\ref{P1}) is an intractable non-convex optimization, mainly due to the coupling between the variables ${\bf p}^{\rm IR}$ and ${\bf x}$ in constraints (\ref{P1-b}) and (\ref{P1-c}), along with the non-convex (\ref{P1-e}) and (\ref{P1-g}).

\section{The Proposed Algorithm for Problem (\ref{P1})}
The variables $\bf x$ and ${\bf p}^{\rm IR}$ are coupled in (\ref{P1-a})-(\ref{P1-c}), which motivates the following two-step method to decouple problem (\ref{P1}). First, given the power allocation variable set ${\bf p}^{\rm IR}$, the position variable set ${\bf x}$ is optimized. After that, the solved ${\bf x}$ is employed to find the solution for ${\bf p}^{\rm IR}$. Finally, two subproblems are alternately solved until convergence.

\subsection{Power Allocation}
{Given the variable ${\bf x}$, the decoding order of the SIC technology can be directly obtained by computing ${{{\big| {\sum\nolimits_{m = 1}^M {h_{{\rm I},{i},m}}} \big|}^2}}$}. Thus, problem (\ref{P1}) can be reformulated as
\begin{subequations}\label{P3}
\begin{eqnarray}
&\mathop {\max }\limits_{{{\bf p}^{\rm IR}}} &\sum\nolimits_{j = 1}^{K_E} {{\sum\nolimits_{i = 1}^{K_I} {\left( {p_{i}^{\rm IR}{{\left| {\sum\nolimits_{m = 1}^M {h_{{\rm E},j,m}}} \right|}^2}} \right)} } } ~~~ \label{P3-a}\\
&{\rm s.t.}&\left(\rm \ref{P1-b} \right), \left(\rm \ref{P1-c} \right), \left(\rm \ref{P1-d} \right).
\end{eqnarray}
\end{subequations}

Problem (\ref{P3}) remains a intractable problem due to the non-convex constraint (\ref{P1-b}). By carrying out an equivalent substitution method on the constraint (\ref{P1-b}), problem (\ref{P3}) can be transformed into
\begin{subequations}\label{P3.1}
\begin{eqnarray}
&\mathop {\max }\limits_{{{\bf p}^{\rm IR}}} &\sum\nolimits_{j = 1}^{K_E} {{\sum\nolimits_{i = 1}^{K_I} {\left( {p_{i}^{\rm IR}{{\left| {\sum\nolimits_{m = 1}^M {h_{{\rm E},j,m}}} \right|}^2}} \right)} } } \label{P3-a}\\
&{\rm s.t.}&{p_{i}^{\rm IR}{{\left| {\sum\limits_{m = 1}^M {h_{{\rm I},i,m}}} \right|}^2} - {\gamma _{\min }} \sum\limits_{l = i+1}^{K_I} {p_{l}^{\rm IR}{{\left| {\sum\limits_{m = 1}^M {h_{{\rm I},i,m}}} \right|}^2}}}    \notag\\
&& \ge {\gamma _{\min }}\sigma _i^2,\forall i \in {\cal K_I},\\
&&\left(\rm \ref{P1-c} \right), \left(\rm \ref{P1-d} \right).
\end{eqnarray}
\end{subequations}

The objective function in problem (\ref{P3.1}) is an affine function with respect to ${\bf p}^{\rm IR}$, and all the constraints are convex constraints. Therefore, it can be efficiently solved via standard convex programming techniques, such as the interior-point method \cite{boyd2004convex}.

\subsection{Positions Optimization}
Obtained the variable ${\bf p}^{\rm IR}$ from problem (\ref{P3.1}), problem (\ref{P1}) can be reformulated as
\begin{subequations}\label{P4}
\begin{eqnarray}
&\mathop {\max }\limits_{{\bf x}} &\sum\nolimits_{j = 1}^{K_E} {{\sum\nolimits_{i = 1}^{K_I} {\left( {p_{i,m}^{\rm IR}{{\left| {\sum\nolimits_{m = 1}^M {h_{{\rm E},j,m}}} \right|}^2}} \right)} } }~~~ \label{P4-a}\\
&{\rm s.t.}&\left(\rm \ref{P1-b} \right), \left(\rm \ref{P1-c} \right),\left(\rm \ref{P1-e}\right),\left(\rm \ref{P1-f}\right),\left(\rm \ref{P1-g}\right).
\end{eqnarray}
\end{subequations}

Problem (\ref{P4}) remains difficult to handle due to the highly coupled variable {\bf x} in the objective function and constraints (\ref{P1-b}) and (\ref{P1-c}). For this challenging problem, we propose two effective algorithms. Specifically, an element-wise algorithm is proposed to derive high-performance solutions, and a LDW-PSO algorithm is proposed to meet the low-complexity requirements.

\subsubsection{Element-wise algorithm}
Each element in the variable $\bf x$ is solved separately. Therefore, given the other elements except $x_k^{\rm PA}$, the new subproblem can be formulated as
\begin{subequations}\label{P5}
\begin{eqnarray}
&\mathop {\max }\limits_{{x_m^{\rm PA}}} &\sum\nolimits_{j = 1}^{K_E} {{\sum\nolimits_{i = 1}^{K_I} {\left( {p_{i}^{\rm IR}{{\left| {\sum\nolimits_{m = 1}^M {h_{{\rm E},j,m}}} \right|}^2}} \right)} } }~~~ \label{P4-a}\\
&{\rm s.t.}&\left(\rm \ref{P1-b} \right), \left(\rm \ref{P1-c} \right),\left(\rm \ref{P1-e}\right),\left(\rm \ref{P1-f}\right),\left(\rm \ref{P1-g}\right).
\end{eqnarray}
\end{subequations}

We can directly employ the one-dimensional search method to search the optimal position of $x_m^{\rm PA}$ within $\left[0,D_w\right]$, provided that the constraints are satisfied. Then, we can successively solve the remaining elements in ${\bf x}$ by utilizing the same method. Noted that the optimal positions of the PAs can be obtained on the premise of discrete position coordinates. In addition, when the position coordinates of the PAs are continuously valued, the one-dimensional search method can still be used to find a near-optimal solution, and its accuracy increases as the search step size decreases. The details of the proposed element-wise algorithm are shown in {\bf Algorithm \ref{alg_ew}}.
\begin{algorithm}[!t]
\caption{Element-wise Algorithm for Solving (\ref{P4})}
\label{penalty}
\begin{algorithmic}[1]
\STATE \small initialize the variable ${\bf x}$, and number of discrete positions $D$
\STATE \small \textbf{repeat}
\STATE \small \hspace{0.3cm}\textbf{for} $m \in {\cal M}$ \textbf{do}
\STATE \small \hspace{0.3cm}\hspace{0.3cm}update $x_m^{\rm PA}$ by solving (\ref{P5}) with one-dimensional search
\STATE \small \hspace{0.3cm}\textbf{end for}
\STATE \small \textbf{until} convergence
\end{algorithmic}
\label{alg_ew}
\end{algorithm}

\subsubsection{LDW-PSO algorithm}
PSO method is a heuristic algorithm that simulates the foraging behavior of bird flocks. PSO algorithm possesses a few advantages, including rapid convergence, ease of implementation, low computational complexity, and the ability to efficiently find high-precision solution in complex environments \cite{985692}. The details of the proposed LDW-PSO algorithm are as follows.

{\textit{Firstly, initialize positions and velocities of the particles, and the individual and swarm optimal positions}}. We initialize $P$ particle swarms, where the particles within each swarm correspond to the $M$ elements in the variable ${\bf x}$. Take an arbitrary particle swarm as an example, the particle vector is simply ${\bf x}$. The velocities of the particles are denoted by ${\bf v} = \left\{v_1,...,v_M\right\}$, which affect the positions of the particles in the next iteration. ${\bf p} = \left\{p_1,...,p_M\right\}$ and ${\bf g} = \left\{g_1,...,g_M\right\}$ represent the current individual and swarm optimal positions of the particles, respectively.

{\textit{Secondly, update the positions and velocities of the particles}}. The velocity of the $m$-th particle can be updated by
\begin{flalign}
{v_m^{{t}}} &= \mathop {\underline {w^{t} \times {v_m^{t-1}}} }\limits_{{\rm{memory~term}}}  + \mathop {\underline {{c_1^{t}} \times {\rm{rand(1)}} \times \left( {p_m^{t-1} - {x_m^{t-1}}} \right)} }\limits_{{\rm{individual-cognitive~term}}} \notag\\
&~~~ + \mathop {\underline {{c_2^{t}} \times {\rm{rand(1)}} \times \left( {g_m^{t-1} - {x_m^{t-1}}} \right)} }\limits_{{\rm{swarm-cognitive~term}}}, \label{vk}
\end{flalign}
where $w^{t}$ is the inertia coefficient in the $t$-th iteration. Both $c_1^t$ and $c_2^t$ are weighting factors in the $t$-th iteration. ${\rm rand(1)}$ represents a random number between $0$ and $1$.

\begin{algorithm}[!t]
\caption{LDW-PSO Algorithm for Solving (\ref{P4})}
\label{penalty}
\begin{algorithmic}[1]
\STATE \small initialize the variables ${\bf x}^0$, ${\bf v}^0$, ${\bf p}^0$, and ${\bf g}^0$
\STATE \small \textbf{for} $t \in {\left\{ 1,...,t_{\max} \right\}}$ \textbf{do}
\STATE \small \hspace{0.3cm}update ${\bf v}^t$ by (\ref{vk})-(\ref{c2})
\STATE \small \hspace{0.3cm}update ${\bf x}^t$ by (\ref{xkt})
\STATE \small \hspace{0.3cm}update ${\bf p}^t$ and ${\bf g}^t$ by (\ref{pg})
\STATE \small \textbf{end for}
\STATE \small obtain ${\bf x}$ by (\ref{xk})
\end{algorithmic}
\label{alg_PSO}
\end{algorithm}
As shown in Eg.(\ref{vk}), the memory term indicates that the update of the velocity is influenced by the current velocity. The individual-cognitive term, which represents the updated velocity stems from its own experience. The swarm-cognitive term, which reflects the collaborative cooperation and knowledge sharing among particles, i.e., the optimal position within the swarm guides the velocity of the particle. Noted that the dynamic weighting coefficient can achieve better search results compared to confusing. Therefore, $w^t$, $c_1^t$, and $c_2^t$ can be updated by
\begin{flalign}
{w^{t}} &= {w_{\max }} - \left( {{w_{\max }} - {w_{\min }}} \right) \times {t} \div {t_{\max }}, \label{wt}\\
c_1^t &= c_1^0 \times {\rm rand}\left(1\right), c_2^t = c_2^0 \times {\rm rand}\left(1\right), \label{c2}
\end{flalign}
where $w_{\max}$ and $w_{\min}$ are the upper bound and lower bound on the inertia coefficient, respectively. $c_1^{0}$ and $c_2^{0}$ are the initial value of $c_1$ and $c_2$, respectively. ${\rm rand}(1)$ denotes a random number in the interval $\left[0,1\right]$. $t_{\max}$ is the maximum number of iterations of the algorithm.

The setting of the parameters in (\ref{vk}) has a significant impact on the search behavior. The larger the value of $w$, the stronger the global optimization ability of the algorithm and weaker its local optimization ability. The values of $c_1$ and $c_2$ determine whether the search tends to the individual optimal position or the swarm optimal position. Based on the updated $v_m^t$, the new position $x_m^{t}$ is given by
\begin{flalign}
x_m^{t} = x_m^{t-1} + v_m^{t}. \label{xkt}
\end{flalign}

{\textit{Finally, update the individual optimal position and the swarm optimal position, and determine whether convergence has been achieved}}. $p_m^{t}$ and $g_m^{t}$ can be obtained by the updated $x_m^{t}$. Then, $p_m^{t}$ and $g_m^{t}$ can be updated by
\begin{flalign}
p_m^t = \max \left\{ {p_m^t,p_m^{t - 1}} \right\}, g_m^t = \max \left\{ {g_m^t,g_m^{t - 1}} \right\}. \label{pg}
\end{flalign}

All particles perform iterative solutions simultaneously according to the above process until the maximum number of iterations $t_{\max}$ is reached, and then the relatively optimal solution of $x_m$ can be exported by
\begin{flalign}
x_m = {g_m^t}. \label{xk}
\end{flalign}

Compared with the developed element-wise algorithm, the LDW-PSO algorithm does not need to search all the points in the feasible region. It can obtain a high-quality solution only through a limited number of iterations, which significantly reduces the computational complexity of the algorithm. The details of the proposed LDW-PSO algorithm are shown in {\textbf {Algorithm \ref{alg_PSO}}}. The approximated computational complexities of the element-wise algorithm and the LDW-PSO algorithm can be expressed as ${\cal O}\left( {LDM} \right)$ and ${\cal O}\left( {PM{t_{\max }}} \right)$, respectively, where $L$ denotes the iteration times of the element-wise algorithm. $P$ denotes the number of the particle swarms in the LDW-PSO algorithm. In this paper, we set $P = 10$, $t_{\max} = 300$, and $D = 2^{12} = 4096$. These settings can achieve the high-performance optimization, as verified by the simulation results in the next section. It can be found that the computational complexity of the LDW-PSO algorithm is significantly lower than that of the element-wise algorithm.

Finally, the solution of the SWIPT-PASS can be obtained by alternately solving the above two subproblems until convergence. 

\section{Simulation Rseult}

\begin{figure}[!t]
\centering
\includegraphics[width=2.9in]{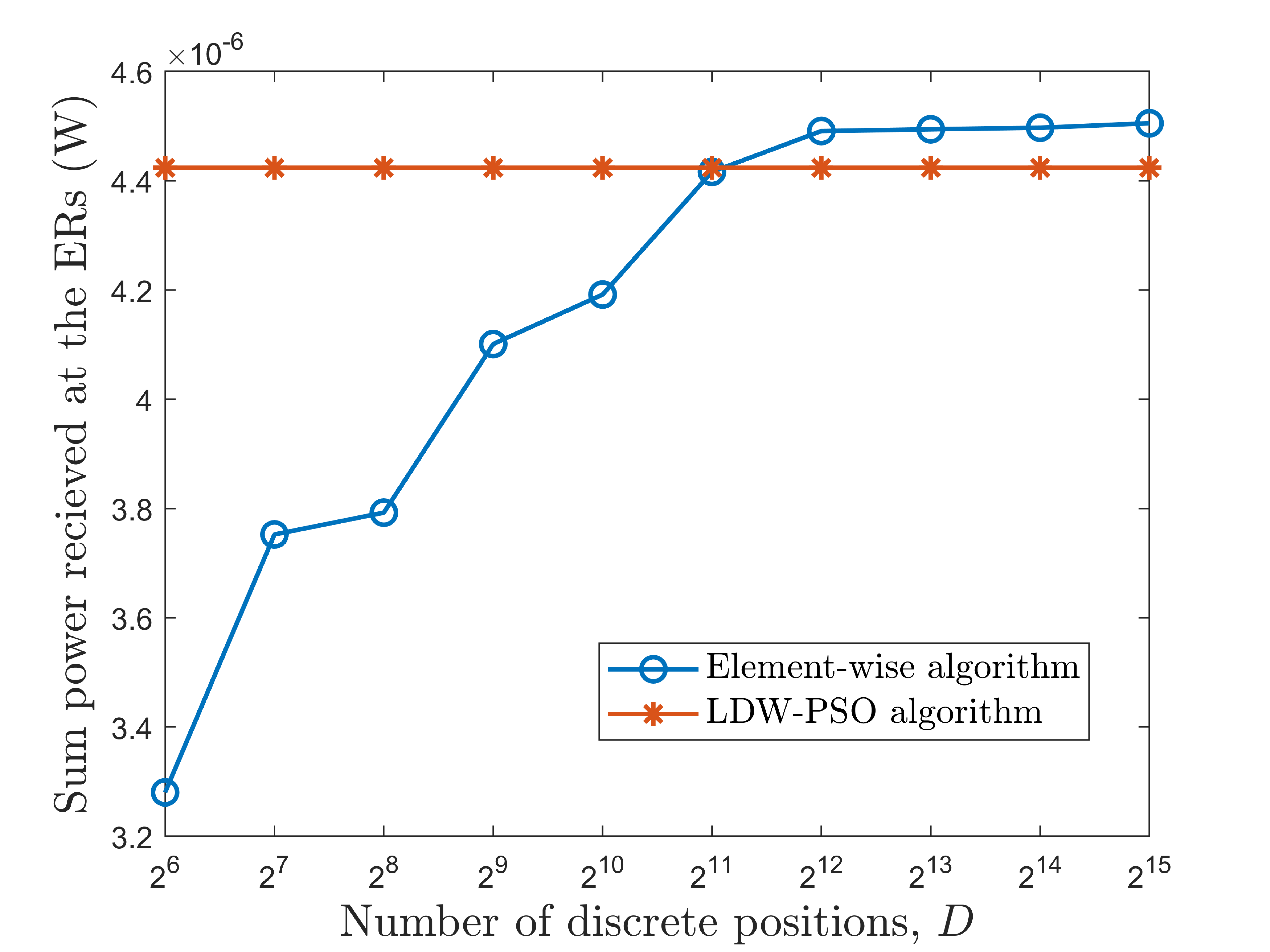}
\caption{Sum power received at the ERs versus $D$.}
\label{D}
\end{figure}

In this section, we demonstrate the performance advantages of the PASS-enabled SWIPT technology by presenting simulation results. First of all, we present the parameter settings based on \cite{10896748} and \cite{10909665} as follows. Let $M = 4$, $K_E = 2$, $K_I = 2$, $P_B = 40~{\rm dBm}$, $\sigma _i^2 = -90~{\rm dBm}, \forall i \in {\cal K_I}$, $f_c = 28~{\rm GHz}$, $n_{\rm neff} = 1.4$, $\lambda = \frac{c}{f_c}$, $c = 3 \times 10^8$, $\Delta  = \frac{\lambda }{2}$, $D_x = 10~{\rm m}$, $D_y = 6~{\rm m}$, $D_w = 10~{\rm m}$, $d = 3~{\rm m}$, $\gamma _{\min} = 15~{\rm dB}$, and $E_{\min} = 0.1~{\mu}{\rm W}$. To comprehensively evaluate the system performance, the benchmark configurations are defined as follows: 1) \textbf{LDW-PSO algorithm:} the proposed LDW-PSO algorithm in this paper; 2) \textbf{Element-wise algorithm:} the proposed element-wise algorithm in this paper; 3) \textbf{MIMO system:} In this case, the $M$ antennas are fixed at static positions near the feed point; 4) \textbf{Fixed-antenna system:} In this case, the a single antenna is fixed at static positions at the feed point. To ensure fairness, all antennas in each benchmark are connected to a single radio frequency chain, as this paper focuses on a single waveguide architecture.

\begin{figure}[!t]
\centering
\includegraphics[width=2.9in]{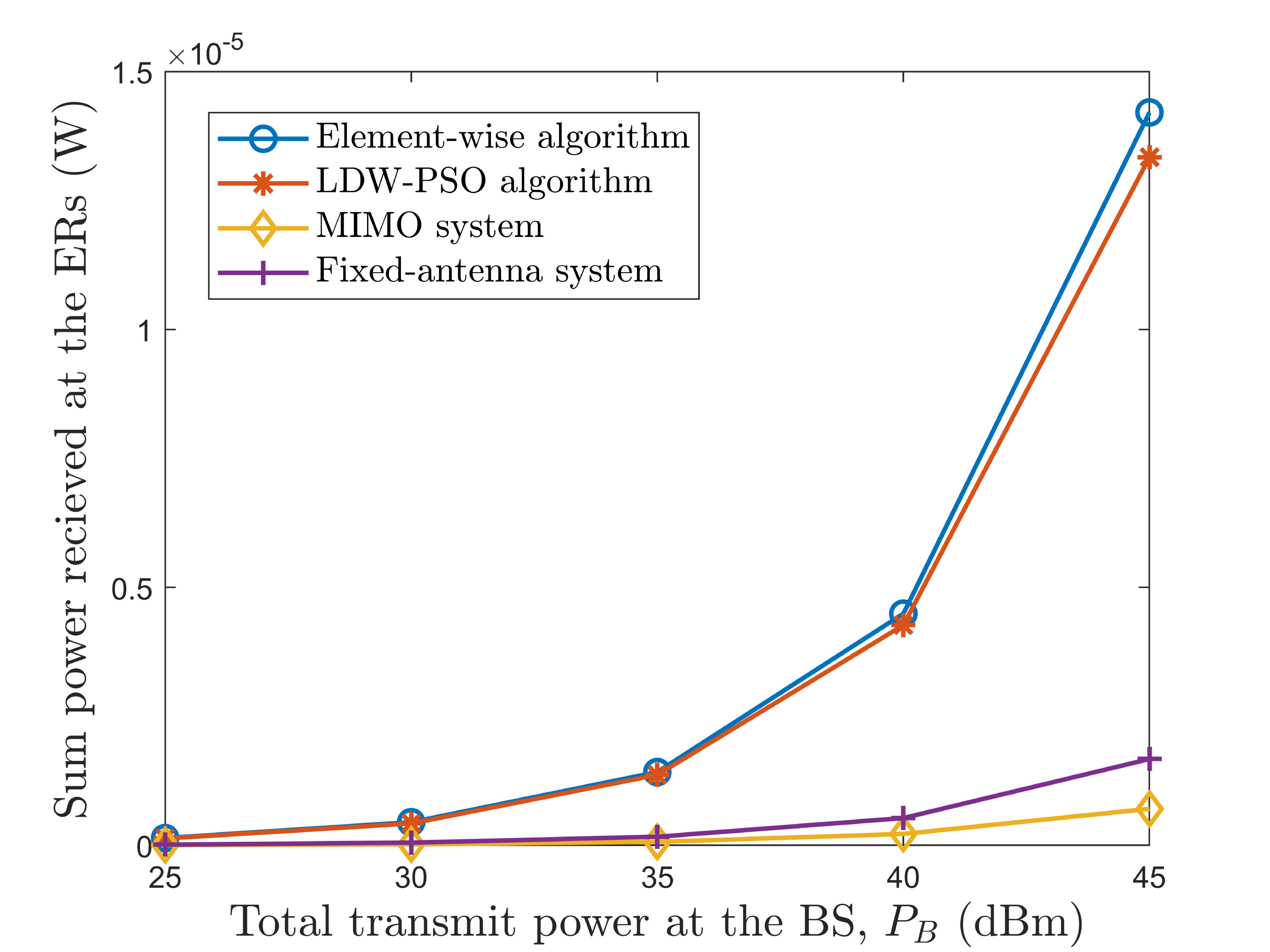}
\caption{Sum power received at the ERs versus $P_B$.}
\label{P}
\end{figure}

\begin{figure}[!t]                                               
\centering
\includegraphics[width=2.9in]{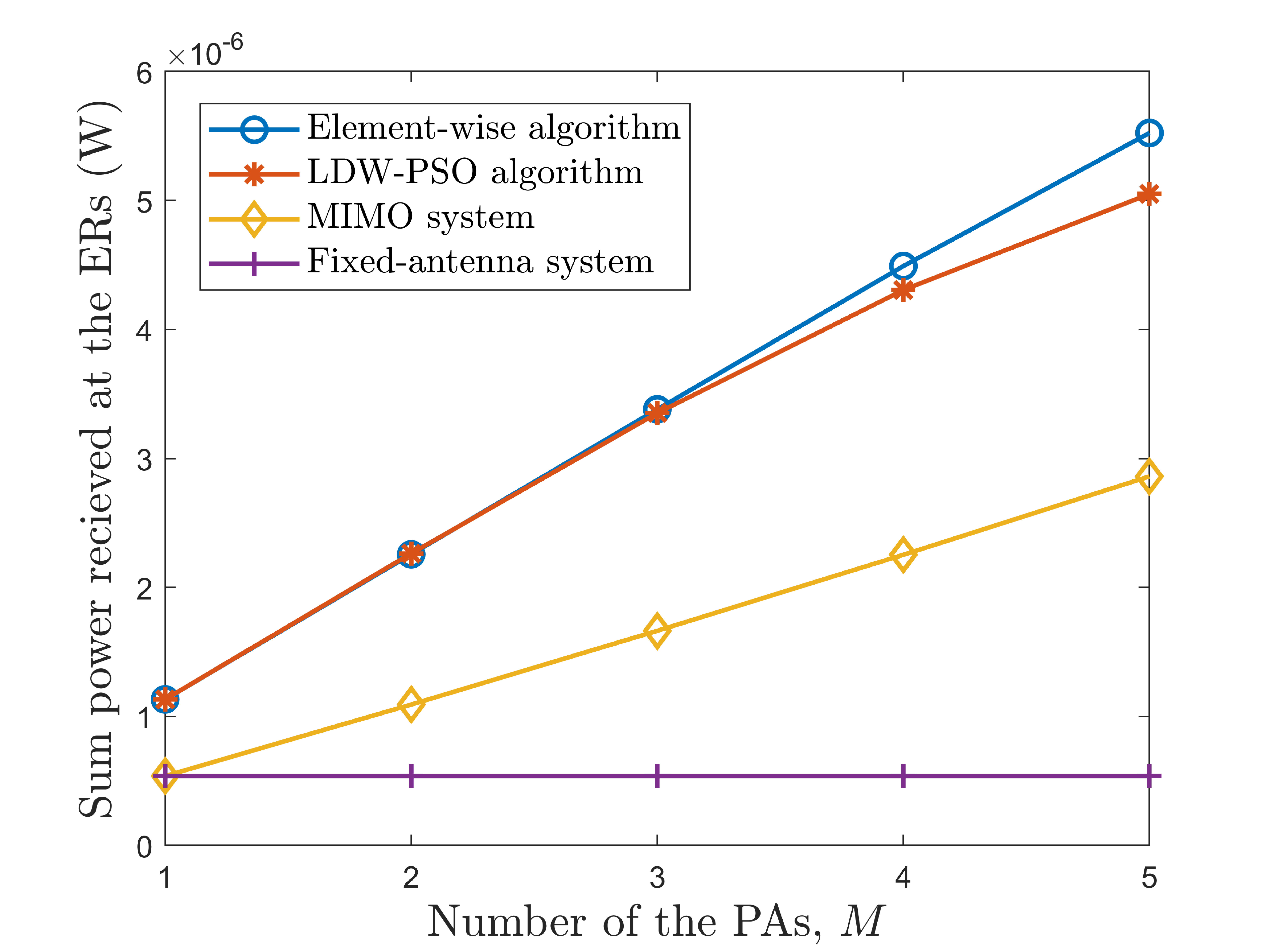}
\caption{Sum power received at the ERs versus $M$.}
\label{M}
\end{figure}
Fig. \ref{D}$-$\ref{M} present the simulation results of the sum power received at the ERs versus the number of discrete positons, the total transmit power at the BS, and the number of the ERs, respectively. The performance of the element-wise algorithm can be considered to be very close to the upper bound due to its search method of discrete values. In these results, the performance of the proposed LDW-PSO algorithm is close to the element-wise algorithm, which verifies the remarkable performance of the LDW-PSO algorithm. Furthermore, we can also find that the wireless power transfer capability of PASS significantly outperforms that of the MIMO system and fixed-antenna system. This is because the PASS can deploy the each PA at an arbitrary position on the waveguide, which yields the simultaneous adjustment of both the large-scale fading and small-scale fading of the channels between the PAs and the IRs/ERs. However, in the MIMO system, transmit beamforming can be realized to adjust the small-scale fading of the channels based on static antennas positions, while the large-scale fading of the channels remains unmitigated. The fixed-antenna system can only passively carry out information transmission and power transfer based on a fixed position.

Fig \ref{D} illustrates the variation in the performance gap between the element-wise algorithm and the LDW-PSO algorithm as the number of discrete positions $D$ increases. The performance of the element-wise algorithm is directly proportional to the variation of $D$. The reason is that an increase in $D$ shortens the search step size, thereby enhancing the search accuracy. When $D$ is greater than $2^{11}$, the performance of the element-wise algorithm is superior to that of the LDW-PSO algorithm, but there is only a small performance gap between the two algorithms. As shown in Fig. \ref{P}, we can find that a greater the transmit power at the BS, leads to a more received power at the ERs will receive. This is because as the total transmit power at the BS increases, the transmit power allocated to the each PA also increases. Therefore, not only the SINR constraints of the IRs more easily satisfied, but also the ERs can received more power. Fig. \ref{M} shows the impact of different number of the PAs on the sum power received at the ERs. As the number of the PAs increases, more sum power received at the ERs can be realized with the BS transmit power kept constant. This is because more PAs can provide higher-precision pinching beamforming for the ERs and IRs. However, it is worth noting that the performance gap between element-wise algorithm and LDW-PSO algorithm grows as the number of PAs increases, because an increase in PA count degrades the convergence accuracy of the LDW-PSO algorithm. Meanwhile, benefiting from the high scalability of the waveguide and the flexible deployment of the PAs along the waveguide, the PASS can significantly improve the power reception gain for the SWIPT system compared to the benchmarks.


\section{Conclusion}
In this letter, we introduced a novel SWIPT enabled by PASS, where adaptive positioning of the PAs allows concurrent reconfiguration of both large-scale and small-scale fading in channels between the BS and IRs/ERs. Firstly, we decoupled the non-convex initial problem into two sub-problems. Specifically, a high-precision element-wise algorithm and a low-complexity LDW-PSO algorithm were proposed for optimizing the position of the PAs. Lastly, an alternating optimization method was adopted to jointly optimize the power allocation and the position optimization. Simulation results demonstrated that the introduced SWIPT-PASS achieves significant performance gain over MIMO system and fixed-antenna system, highlighting the superiority of reconfigurable large-scale fading mitigation in wireless power transfer.

\bibliography{Reference_PA-SWIPT}
\bibliographystyle{IEEEtran}

\end{document}